\documentclass[journal=jacsat,manuscript=article]{achemso}

\usepackage[version=4]{mhchem} 
\usepackage{graphicx}
\usepackage{amsmath,amssymb,amsfonts}
\usepackage{algorithmic}
\usepackage{textcomp}
\usepackage{xcolor}
\usepackage{gensymb}
\usepackage{siunitx}
\usepackage{multirow, makecell}
\usepackage{tabularx}
\usepackage{hyperref} 
\usepackage{ulem}

\author{\href{https://orcid.org/0000-0003-0027-9112}{\includegraphics[scale=0.06]{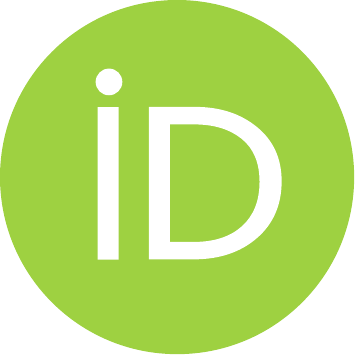}\hspace{1mm}Robin Lef\`evre}}
\email{robin.lefevre@uzh.ch}
\phone{+41 44 63 54630}
\affiliation[University of Zurich]
{Department of Chemistry, University of Zurich, CH-8057 Z\"urich, Switzerland}
\author{\href{https://orcid.org/0000-0003-0027-9112}{\includegraphics[scale=0.06]{orcid.pdf}\hspace{1mm}Fabian O. von Rohr}}
\email{fabian.vonrohr@uzh.ch}
\phone{+41 44 63 54630}
\affiliation[University of Zurich]
{Department of Chemistry, University of Zurich, CH-8057 Z\"urich, Switzerland}

\title{A Heavy-Fermion Zn-deficient \ce{CaBe2Ge2}-Type Phase with Rare Ce-based Ferromagnetism and Large Magnetoresistance}

\begin{document}

\begin{tocentry}

\includegraphics[]{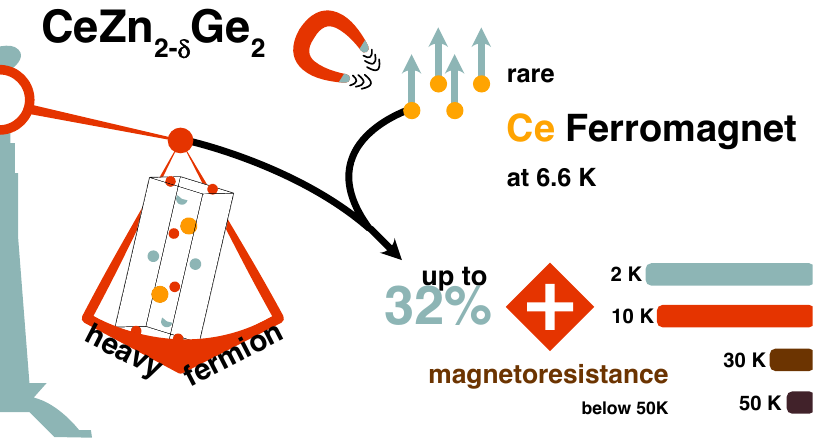}

\end{tocentry}

\begin{abstract}
We report on the \textit{hitherto} unknown compound \ce{CeZn$_{2-\delta}$Ge2} ($\delta$ $\approx$ 0.41). We find this compound to crystallize in a defect version of the well-known \ce{CaBe2Ge2} structure type. The phase forms in a Zn/In flux and with Zn-deficiency on one of its crystallographic sites. We find the compound to display uncommon localized Ce-based (4f$^1$) ferromagnetism with a $T_{\textrm{C}}$ = 6.6 K, a large positive magnetoresistance reaching an MR of approximately 32 \% below $T$ = 10 K, and strongly correlated electrons, as evidenced by a Kadowaki-Woods ratio A/$\rm \gamma ^2$ close to known heavy fermion compounds. The here discovered material is therefore a promising model platform for the investigation of these entangled interacting and potentially competing electronic states paired with complex crystal chemistry.\\
\\
\end{abstract}

\section{Introduction}
The discovery of new quantum materials has proven time and time again to be a crucial stimulus for fundamental research and eventually technological progress.\cite{tokura_emergent_2017,giustino_2021_2021,maple_new_2009} For the discovery of new materials with strongly correlated electrons, chemical design principles and exploratory chemical syntheses are essential for advancements in this field.\cite{jansen_concept_2002,khoury_new_2019,bao_quasi-two-dimensional_2021,von_rohr_high-pressure_2017} Especially, as the computation -- \textit{e.g.} the implementation in to density functional calculation -- of electronic correlations remains today a major challenge.\cite{carleo_solving_2017} In particular, heavy fermion compounds that exhibit transitions to emergent collective quantum states are of special interest, due to the entanglement of these different interacting and potentially competing electronic states.\cite{fulde_strongly_2006,georges_strong_2013} These materials are for instance promising candidates for the observation and study of quantum phase transitions, which occur at zero temperature and must be fundamentally different in nature from the conventional, thermal phase transitions.\cite{shen_strange-metal_2020,huang_quantum_2016,brando_metallic_2016,lohneysen_fermi-liquid_2007}   

More than 240 compounds crystallizing in the \ce{CaBe2Ge2} structure are known.\cite{zheng_donor-acceptor_1986,venturini_contribution_1989} This structure is closely related to the \ce{ThCr2Si2} structure, which, with more than 2400 known phases, is considered the most common intermetallic structure-type.\cite{shatruk_thcr2si2_2019} \ce{XY2Z2} compounds crystallize in the two structure-types \ce{ThCr2Si2} and \ce{CaBe2Ge2} when they exhibit the \textit{I}4/\textit{mmm} or \textit{P}4/\textit{nmm} space-groups, respectively. If one takes the succession of X, Y, Z as equivalent atomic positions forming the square lattices perpendicular to \textit{c}, ..[XZYZXZYZX].. is the sequence in \ce{ThCr2Si2}, which then turns into ..[XY'ZY'XZ'YZ'X].. in \ce{CaBe2Ge2}, with Y/Y' and Z/Z' being non-equivalent atomic positions. These structures are known to be excellent hosts for heavy fermionic properties. In these structures, the magnetic 4f ions occupy the corners and the center of the unit cell. Next-nearest neighbor in-place distances between 4f ions and second shortest out-of-plane distances are almost identical. At the same time the structure can be highly layered when nonexistent inter-layer chemical bonds occur, with \ce{Y2Z2}-layers that may exhibit anisotropic metallic electrical conductivity. The most prominent heavy fermion compound in these structures is the compound \ce{CeCu2Si2}, which was one of the first experimental realizations of heavy fermion behavior and the first heavy fermion superconductor.\cite{steglich_superconductivity_1979,mathur_magnetically_1998} Especially noteworthy are also the exotic physical properties of \ce{ThCr2Si2}-type compound \ce{URu2Si2}, which is an 17 K antiferromagnet and undergoes a transition to a superconducting state at $T_{\rm c} \approx$ 1 K.\cite{stockert_superconductivity_2012,palstra_superconducting_1985}

Cerium intermetallic compounds have attracted wide interest due to their crystal chemistry and their correlated electronic properties, such as heavy fermions, superconductivity, magnetic ordering, Ce$^{3+}$/Ce$^{4+}$ valence fluctuations, and combinations thereof.\cite{steglich_superconductivity_1979,nagao_superconductivity_2016,prokleska_magnetism_2015,kratochvilova_coexistence_2015,matusiak_anomalous_2013} Additionally, Ce$^{3+}$ compounds with a localized 4f$^1$ electron are prompt to form a non-magnetic Kondo singlet\cite{kondo_resistance_1964} while Ruderman-Kittel-Kasuya-Yosida (RKKY) interactions will activate antiferromagnetic long-range magnetic order.\cite{kasuya_theory_1956,ruderman_indirect_1954} Hence, if they do show magnetic order, cerium compounds commonly display antiferromagnetic ordering.\cite{pottgen_cerium_2015,rossat-mignod_magnetic_1983} Ferromagnetism in Ce-compounds has been of great interest, especially as the localized 4f$^1$ electron in the Ce$^{\rm 3+}$ state has been found to more often display antiferromagnetic ordering, rather than ferromagnetic ordering.\cite{brando_metallic_2016,hafner_kondo_2019,stewart_non-fermi-liquid_2001} 

Ce-based ferromagnets, in which the ferromagnetism arises likely from the localized moments of the 4f$^1$ electrons include \ce{CeRuPO}, \ce{CeRu2Ge2}, \ce{CePdSb}, \ce{CeAuGe}, and \ce{CeRu2Al2B}.\cite{krellner_cerupo_2007,matusiak_anomalous_2013,tran_ferromagnetism_2014,malik_cepdsb_1991,pottgen_ferromagnetic_1996,tursina_cepd2al8_2018,givord_ferromagnetism_2007,baumbach_ceru_2_2b_2012}  The Ce-compound in which cerium crystallizes with non-magnetic elements to form a different structure and showing the highest ferromagnetic transition temperature is \ce{CeRh3B2}. The ferromagnetism in this compound has been identified to be itinerant in nature.\cite{yaouanc_cerh3b2_1998,kishimoto_11b_2004}  
In the Ce-Ge-Zn ternary system, so far 5 compounds have been reported, namely \ce{Ce2Zn6Ge3}, \ce{Ce2Zn3Ge6}, \ce{CeZnGe}, \ce{Ce7Zn_{21.30}Ge_{1.7}} and \ce{Ce2Zn$_{0.91}$Ge6}. \ce{Ce2Zn3Ge6} is an antiferromagnet at low temperatures with a N\'eel temperature below 3 K, \cite{salvador_intermetallic_2005} while \ce{Ce2Zn6Ge3} is a metal with magnetic ordering at 7.2 K.\cite{grytsiv_novel_2003} 
Here, we report on a new compound in the \ce{CaBe2Ge2} family. The structure of \ce{CeZn$_{2-\delta}$Ge2} with $\delta$ $\approx$ 0.41 has been determined by means of single-crystal X-ray and powder neutron diffraction. We find this phase to form in a Zn/In flux, but nevertheless resulting in a Zn-deficient phase. \ce{CeZn$_{2-\delta}$Ge2} crystals are metallic with a ferromagnetic transition at $T_{\rm C}$ $\approx$ 6.6 K. Below 50 K, a magnetoresistance arises with the field, a temperature way higher than the start of the ordering of the spins. Below 10 K, a positive magnetoresistance up to 32\% is observed. From specific heat and resistivity measurements, we conclude that \ce{CeZn$_{2-\delta}$Ge2} is a new heavy fermion compound.

\section{Results and discussion}
\subsection{Crystallography of \ce{CeZn$_{2-\delta}$Ge2}}

\begin{figure*}
\centering
\includegraphics[width=\textwidth]{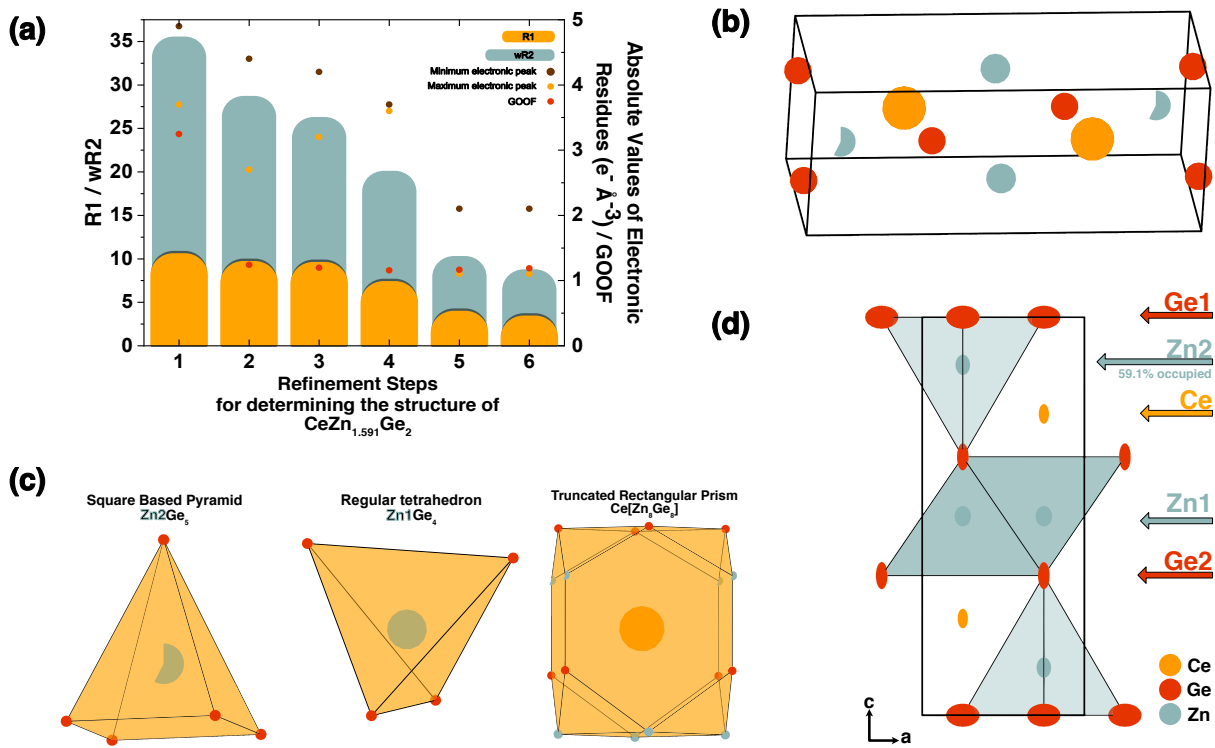}
\caption{(a) Details on the structural resolution and refinement. The converged values for reliability factors R1, wR2 and GOOF as well as min/max residual electronic densities for each steps of the refinement are represented. Step description is mentioned in the text. (b) 3D-representation of the unit cell. (c) Coordination polyhedra of Zn2, Zn1 and Ce atoms. (d) [010] oriented representation of the unit cell with anisotropic displacement parameters represented at 80\% probability.}
\label{fig:fig1}
\end{figure*}

\ce{CeZn$_{2-\delta}$Ge2} crystals were synthesized in a Zn/In flux, following the synthesis route described in the methods section. The crystals have a thin platelet-like shape with a metallic luster and sizes up to 5 $\times$ 5 $\times$ 0.4 mm$^3$, visibly layered and brittle. The structure of the previously unknown compound was determined by means of SXRD. The final cell parameters in the tetragonal crystal system are based on 588 reflections and refined to $a$ = 4.20635(16) $\si{\angstrom}$ and $c$ = 10.3338(7) $\si{\angstrom}$. Spherical absorption and empirical absorption corrections using spherical harmonics were applied to the reflections. The compound crystallizes in the space-group \textit{P}4/\textit{nmm}. It is then isostructural to the \ce{CaBe2Ge2} structure. The structure was solved and refined over a series of 6 steps, where the subsequent improvement in the agreement factors R1 and wR2 strongly indicates the validity of the obtained structural solution (see fig.\ref{fig:fig1}. As shown in figure \ref{fig:fig1}(a): \textbf{(Step 1-3)} From initial \ce{CeZn2Ge2} stiochiometry, correctly assigned Zn/Ge according to reference \citenum{grytsiv_euzn2si2_2002} and usual weighting scheme and extinction coefficient, we observe a large improvement of the agreement factors R1 and wR2. \textbf{(Step 4)} was initialized by freeing all site occupancies. As a result, we found that almost all sites are fully occupied, with the Zn2 site only close to being half-occupied. Resulting in the Zn2 site being fixed at 50\% while all the other sites were fixed at a 100\% occupancy. In \textbf{Step 5} the anisotropic displacement parameter were freed for all atoms. In a final \textbf{Step 6}, the structural refinement was concluded by freeing the occupancy of the Zn2 site, leading to a 59.1 $\pm$ 1.1\% occupancy. Upon application of the previously mentioned steps, the structure refinement lead to excellent reliability factors for a defect structure of R1 = 2.96\% and wR2 = 7.06\% demonstrating the veracity of the refinement. Details of the data collection and structure refinement are given in table 1. In table 2 the atomic positions and their respective ADPs and in table 3 the interatomic distances are listed. The final refined stoichiometry is \ce{CeZn$_{1.591(11)}$Ge2}. Multiple probes on different crystals led to equivalent zinc content ranging from 1.569 to 1.591, the structure seems to crystallize with an average of 1.58 zinc. This stoichiometry is confirmed by our EDS analyses in which we find a zinc deficiency leading to an average composition of Ce:22.76$\pm$0.62\% (exp: 21.8\%), Zn:35.41$\pm$1.03\% (exp: 34.7\%) and Ge:41.83$\pm$0.97\% (exp: 43.5\%) (see Supporting Information SFig. 2 and STable 1).

The structure has five crystallographic sites, one occupied by cerium, two by zinc and two by germanium. All but one sites are fully occupied. A zinc deficiency is observed on the site located at [3/4,3/4,0.1199(4)], reaching a final occupancy of 59.1\% for Zn2. \ce{CeZn$_{1.591(11)}$Ge2} is isostructural to \ce{CaBe2Ge2}. The structure is based on a network of bonded Zn/Ge atoms, surely altered by the random distribution of the Zn2. This atom distribution is represented in figure \ref{fig:fig1}(b). The network is based on \ce{Zn{2}Ge5} squared base pyramids (see figure \ref{fig:fig1}(c)) and \ce{Zn{1}Ge4} tetrahedra (see figure \ref{fig:fig1}(c)). Zn1 tetrahedra form a layer along the ab-plane, as in \ce{ThCr2Si2}, they are connected by their edges and form a centered square lattice. Distances for Zn1 tetrahedra are found to be d(Ge1-Zn1) = 2.6065(14) $\si{\angstrom}$. The \ce{Zn{2}Ge5} pyramids are based on a double layer connected by four base corners and forming a centered square pattern. This corresponds to a mirroring pointing upwards and downwards, while one of the layer is shifted by ($\frac{1}{2}$,$\frac{1}{2}$) on x and y (\textit{i.e.} \textit{n} glide plane), creating the centered square pattern. Distances for these pyramids are d(Zn2-Ge1) = 2.389(5) $\si{\angstrom}$ away from and d(Zn2-Ge1) = 2.441(2) $\si{\angstrom}$ towards the mirror plane. Only three compounds are found in the \ce{AZn2Ge2} ternary systems within the families \ce{ThCr2Si2} and \ce{CaBe2Ge2}, all showing divalent cation A (\textit{i.e}. A = Eu, Sr or Ca).\cite{grytsiv_euzn2si2_2002,dorrscheidt_notizen_1976,eisenmann_neue_1970} In all compounds, the Zn-Ge distances are close to the one found in our \ce{CeZn$_{1.591(11)}$Ge2}, with typical distances being 2.506 \si{\angstrom} for the pyramidal \ce{ZnGe5}, 2.579 \si{\angstrom} and 2.610 \si{\angstrom} in the layer of \ce{ZnGe4} tetrahedra (in \ce{EuZn2Ge2}).\cite{grytsiv_euzn2si2_2002}

Ce sits in encaged tunnels formed by those polyhedra, in a truncated rectangular prism with main length and width to be 2.6565(7) $\si{\angstrom}$ and 2.1031(2) $\si{\angstrom}$ (see figure \ref{fig:fig1}(c)). Distances Ce to network are found to be d(Ce-Ge1) = 3.1769(9) $\si{\angstrom}$, d(Ce-Ge2) = 3.2755(7) $\si{\angstrom}$, d(Ce-Zn1) = 3.3877(7) $\si{\angstrom}$ and d(Ce-Zn2) = 3.2351(17) $\si{\angstrom}$. These distances are in agreement with the less ionic europium-II related-compound \ce{EuZn2Ge2} showing distances Eu to Zn/Ge ranging from 3.283 $\si{\angstrom}$ to 3.457 $\si{\angstrom}$.\cite{grytsiv_euzn2si2_2002} Additionally, Ce$^{3+}$  and Ca$^{2+}$ have equivalent ionic radii, and the distances from calcium to germanium sites in \ce{CaZn2Ge2} are found to be 3.224 $\si{\angstrom}$. \cite{eisenmann_neue_1970} 

The Ce network is arranged in a body centered tetragonal prism with the size of the cell unit. Direct magnetic interaction is formed between the nearest-neighbor atoms with a minimal d(Ce-Ce) = 4.2063(2) $\si{\angstrom}$), a direct Ce-Ce distance being relatively large. For reference, the antiferromagnet \ce{CeRu2As2} and ferromagnet \ce{CePd2P2} show also Ce-Ce distances close to 4.20 $\si{\angstrom}$, the two showing the same square lattice arrangement of cerium atoms.\cite{cheng_synthesis_2019,tran_ferromagnetism_2014}

All anisotropic displacement parameters are low. Fig. \ref{fig:fig1}(d) represents the unit cell along \textit{b} with anisotropic displacement parameters being displayed. The U$_{11}$ and U$_{33}$ displacement parameters of Ge2 and Ge1, respectively, are slightly larger than the others. Ge2 will move in the \textit{ab}-plane and Ge1 along \textit{c} to compensate the deficiency of Zn2 in certain unit cells. 

The purity of the sample was first checked using PXRD, Rietveld refinement leads to small impurity of indium and zinc fluxes, with main phase \ce{CeZn_{2-\delta}Ge2} at about 97\% (see SFig. 1). With low indium content, \textit{i.e. small absorption} powder neutron diffraction was performed in order to assess the position attributed to germanium and zinc in the structure, as well as zinc deficiency. On SFig. 3, the Rietveld refinement on neutron data with final agreement factors of R1 = 3.80\% and wR2 = 4.85\% and low residual electronic densities of +0.0455 and -0.0503 $e^{-}$ $\si{\angstrom}^{-3}$ confirms the previous structural model obtained from single X-ray diffraction, no mixing between zinc and germanium is observed. Additionally, a Zn2 occupancy of 57\% is found, a value almost identical to the SXRD refinement. 

Non-stoichiometry is known in both the \ce{ThCr2Si2} and \ce{CaBe2Ge2} structural families, \textit{e.g.} \ce{CeNi_{1.26}Sb2}, \ce{CaCu_{1.75}As2} or \ce{EuNi_{1.53}Sb2}.\cite{hofmann_structural_1984,schafer_ceni1-xsb1ybi1-y_2014,pilchowski_ternare_1990} Here, for CeZn$_{1-\delta}$Ge$_2$ the Zn deficiency is found while Zn is in excess. Hence, it seems to be a universal property of this compound and not a variable parameter. Moreover, preliminary powder neutron diffraction data confirms the absence of mix-atomic sites and the zinc deficiency.

Bobev \textit{et al.} proposed that non-stoichiometry allow stabilization of the electronic structure by removing excess of valence electrons.\cite{bobev_nickel_2009} For our particular case, further studies will be needed to confirm or contradict this hypothesis.

\begin{table*}
\centering
\caption{Crystallographic data for single-crystal \ce{CeZn$_{1.591(11)}$Ge2}.}
\label{table:table1}
 \resizebox{0.5\textwidth}{!}{
 \begin{tabularx}{0.7\columnwidth}{p{5cm} p{6cm}}
  \multicolumn{2}{l}{\textbf{Physical, crystallographic, and analytical data}}\\
  \hline 
 Formula & \ce{CeZn$_{1.591(11)}$Ge2} \\
 Structure type & \ce{CaBe2Ge2}\\
 Mol. wt. (g mol$^{-1}$) & 389.24 \\
 Cryst. syst. & Tetragonal \\
 Space group & \textit{P}4/\textit{nmm} (129)\\
 a ($\si{\angstrom}$) & 4.20635(16)\\
 c ($\si{\angstrom}$) & 10.3338(7)\\
V ($\si{\angstrom}^3$) & 182.840(18) \\
Z & 2 \\
Calculated density (g cm$^{-3}$)& 7.070 \\
Temperature (K) & 160.0(1) \\
  \multirow{2}{*}{Diffractometer}& XtaLAB Synergy  \\
 & Dualflex, Pilatus 200K \\
 Radiation & Cu K$\alpha_1$ (1.54184 $\si{\angstrom}$)\\
 Crystal color & Metallic black \\
 Crystal description & Piece from a larger crystal \\
 Crystal size (mm$^3$) & 0.039 $\times$ 0.017 $\times$ 0.011 \\
 \makecell[l]{Linear absorption \\ coefficient (cm$^{-1}$)} & 1226  \\
 Scan mode & $\omega$ \\
 Recording range $\theta$ ($^{\circ}$) & 4.278 --- 71.149\\
 \multirow{3}{*}{hkl range} & $-5 \leq h \leq 5$ \\
  & $-5 \leq k \leq 5 $ \\  
  & $-9 \leq l \leq 12 $  \\  
  Nb. of measured reflections & 908 \\ [4ex] 
  
 \multicolumn{2}{l}{\textbf{Data reduction}} \\
 Completeness (\%) & 100 \\
  No. of independent reflections & 137 \\
  Rint (\%) & 4.22 \\
 \multirow{2}{*}{Absorption correction}& Analytical using spherical \\
 & harmonics and frame scaling \\
 \makecell[l]{Independent reflections \\ with I $\geq$ 2.0$\sigma$ } & 125 \\ [4ex]
 
  \multicolumn{2}{l}{\textbf{Refinement}}\\
 R1 (obs / all) (\%)  & 2.96 / 3.18  \\
 wR2 (obs / all) (\%) & 7.06 / 7.20 \\
 GOF & 1.177 \\
 No. of refined parameters & 16 \\
 \makecell[l]{Difference Fourier\\ residues ($e^{-}$ $\si{\angstrom}^{-3}$) }& -2.095 --- +1.274  \\
  \hline
\end{tabularx}}
\end{table*}

\begin{table*}
\centering
\caption{Refined coordinates, Wyckoff positions, occupancies, atomic displacement parameters (ADPs), and their estimated standard deviations for single-crystal \ce{CeZn$_{1.591(11)}$Ge2}.}
\label{table:table2}
 \resizebox{\textwidth}{!}{
 \begin{tabularx}{\columnwidth}{p{1cm}p{1.2cm}Xp{0.8cm}p{0.8cm}XXXXX}
 \hline 
 Atom & \makecell[l]{Wyckoff \\ position} & Occ. & x & y & z & $U_{iso}$ ($\si{\angstrom}$ $^2$) & U$_{11}$ & U$_{33}$ \\
 \hline 
 Ce & 2\textit{c} & 1 & 1/4 & 1/4 & 0.24300(9) & 0.0085(5) & 0.0049(5) & 0.0155(7)\\
 Ge1 & 2\textit{c} & 1 & 3/4 & 3/4 & 0.3510(2) & 0.0134(7) & 0.0058(7) & 0.0286(13)\\
 Ge2 & 2\textit{a} & 1 & 1/4 & 3/4 & 0 & 0.0340(8) & 0.0407(12) & 0.018(2)\\
 Zn1 & 2\textit{b} & 1 & 1/4 & 3/4 & 1/2 & 0.0125(7) & 0.0105(9) & 0.0206(14)\\
 Zn2 & 2\textit{c} & 0.591(11) & 3/4 & 3/4 & 0.1199(4) & 0.0120(11) & 0.0089(17) & 0.018(2)\\
 \hline
  \multicolumn{9}{l}{U$_{11}$ $=$ U$_{22}$ and U$_{12}$, U$_{13}$, U$_{23}$ $=$ 0 due to symmetry constraints.}
\end{tabularx}
}
\end{table*}

\begin{table*}
\centering
\caption{Principal interatomic distances and their standard deviations in \ce{CeZn$_{1.591(11)}$Ge2} given by the single-crystal study.}
\label{table:table3}
\resizebox{\textwidth}{!}{
 \begin{tabularx}{\columnwidth}{p{1cm}p{1cm}XXp{1cm}p{1cm}XX}
 \hline
 Atom1 & Atom2 & Distance (\AA)& Multiplicity & Atom1 & Atom2 & Distance (\AA) & Multiplicity \\
 \\
  Ce & Ge1 & 3.1769(9) & x 4 &  Ce & Ge2 & 3.2755(7) & x 4\\
 Ce & Zn2 & 3.2351(17) & x 4 & Ce & Ce & 4.2063(2) & \\
  Ge1 & Zn1 & 2.6065(14) & x 4 & Ge1 & Zn2 & 2.389(5) & x 1\\
 Ge2 & Zn2 & 2.441(2) & x 1 & & & &\\
 \hline
\end{tabularx}
}
\end{table*}

\subsection{Magnetic properties}

\begin{figure*}
\centering
\includegraphics[width=\textwidth]{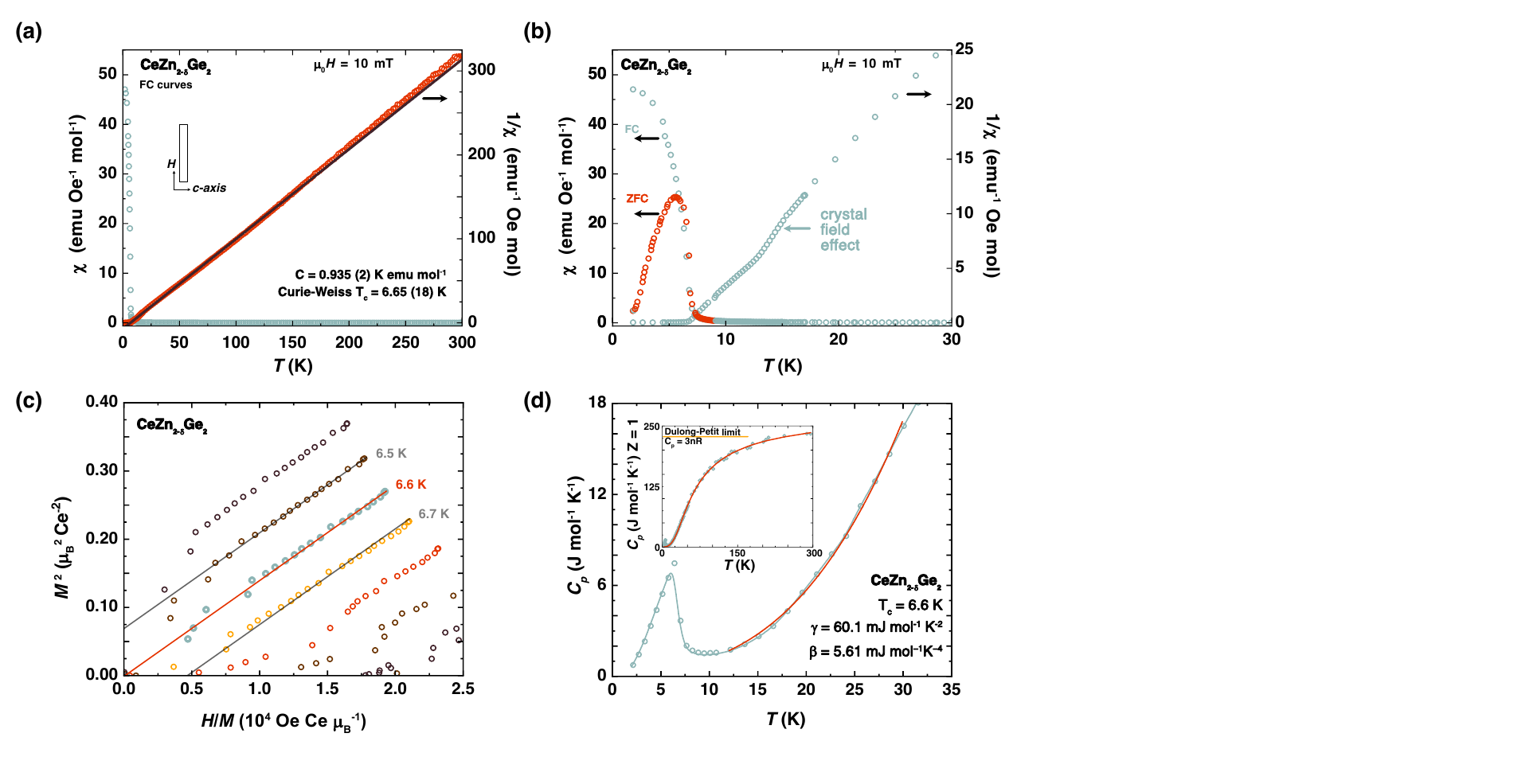}
\caption{(a)-left axis: temperature dependence of the in-\textit{ab}-plane magnetic susceptibility between $T$ = 300 K and 1.8 K, the FC curve is represented. -right axis: inverse of the magnetic susceptibility with the Curie-Weiss fit from 16 K to 175 K. (b) -left axis: the temperature range is zoomed from 1.8 to 30 K for both the susceptibilities, \textit{i.e.} ZFC and FC curves. -right axis: shows the inverse of the FC curve.(c) Arrott plot constructed from temperature field dependence of the in-\textit{ab}-plane magnetization for \ce{CeZn$_{2-\delta}$Ge2}, from 6.4 K to 7 K up to 0.1 T. $T_{\textrm{C}}$ is found at 6.6 K. (d) Heat capacity between 35 K and 2 K and its fit above the transition. -inset: C$\rm {_P}$ from 2 K to 300 K with high-temperature values reaching the Dulong-Petit limit. A 13 K to 300 K fit is shown using the Debye-Einstein fitting, see eq. \ref{debye}.}
\label{fig:fig2}
\end{figure*}

Zero-field cooled and field-cooled in-\textit{ab}-plane susceptibilities $\chi$ under $\mu_0 H$ = 1 T are presented in supplementary material SFig. 4 and are identical. A field of 1 T is strong enough to forbid spin glass effect. In fig. \ref{fig:fig2}(a), the temperature-dependent field-cooled susceptibility of \ce{CeZn$_{2-\delta}$Ge2} measured in a magnetic field of $\mu_0 H$ = 10 mT is plotted in a temperature range between $T$ = 1.8 K and 300 K. Below 10 K, we observe a sharp uplift of the magnetization. On the FC-curve, a saturation of the moments is reached below the transition, a typical behavior for a ferromagnet. Fig. \ref{fig:fig2}(b) shows a closer look at the inverse of the susceptibility and displays an enhancement of the susceptibility below the crystal field effect temperature. The magnetization above 10 K corresponds to the behavior of a paramagnet, and it is well described by a Curie-Weiss law. A fit was performed on the inverse of the magnetic susceptibility curve from 16 K to 175 K using the equation 
\begin{equation}
\mathrm{1/\chi = \textit{T}/C - \theta_C/C}
\end{equation}
with C being the Curie constant and $\theta_C$ the Curie-Weiss temperature. The Curie constant of C = 0.935(2) emu mol$_{Ce}^{-1}$ and a Curie-Weiss temperature of $\theta_C$ = 6.65(18) K are extracted from the fit. The extracted $\theta_C$ agrees with nearest neighbors ordering ferromagnetically interactions and is very close to the transition temperature obtained from the derivative curve of the susceptibility (see SFig. 5). From the Curie constant, we can extract an effective paramagnetic moment $\rm \mu_{eff}$ = $\sqrt{8C}$ = 2.65 $\rm \mu_B$ Ce$^{-1}$. The estimated $\rm \mu_{eff}$ value is a bit larger, but close to the calculated Ce$^{3+}$ effective moment value of 2.54 $\rm \mu_B$ Ce$^{-1}$, evidencing that the magnetism in this material arises due to the localized Ce$^{3+}$ 4f$^1$ electrons.

The ferromagnetic transition temperature $T_{\textrm{C}}$ was determined from both an Arrott plot \cite{arrott_criterion_1957} and $C_{p}$ specific heat capacity measurements in the vicinity of the transition temperature, see fig. \ref{fig:fig2}(c) and (d). In the corresponding Arrott plot, the transition temperature $T_{\textrm{C}}$ is found for the line without an intercept when plotting $M^2$ vs. $H$ / $M$ for various temperatures. The transition temperature of \ce{CeZn$_{2-\delta}$Ge2} is thereby determined to be $T_{\textrm{C}}$ = 6.6 K. In \ce{CeZn$_{2-\delta}$Ge2}, no other magnetic metal is present, and only cerium magnetic moments lead to this magnetism. Powder x-ray diffraction performed on crushed crystals did not evidence additional magnetic impurities. Ferromagnetism from localized moments is rarely found in Ce-based magnets. Both figures \ref{fig:fig2}(c) and (d) show a magnetic transition at 6.6 K. From using the heat capacity equation in fig \ref{fig:fig2}(d), fits were realized on temperatures higher than the transition (12.5 K to 30 K) using eq. \ref{T3} and a linear regression of C/T \textit{vs.} T$^2$, as shown in fig. \ref{fig:fig2}d and Sfig. 6, respectively. Additionally, in order to  confirm the obtained fitted parameters, the full Debye-Einstein equation was used to fit the temperature range from 13 to 300 K (see eq. \ref{debye}). 

\begin{equation}
\label{T3}
C(T) = \gamma T + \beta T^3
\end{equation}

\begin{equation}
\label{debye}
    C(T) = \gamma T + 9n\mathrm{R} (1-d) (\frac{T}{\Theta_D})^3 {\int_{0}^{\frac{\Theta_D}{T}} \frac{x^4 e^x}{(e^x - 1)^2} \,dx} + 3n\mathrm{R}d (\frac{\Theta_E}{T})^2 \frac{e^{\Theta_E/T}}{(e^{\Theta_E/T} - 1)^2}
\end{equation}

where the first term represent the electronic contribution ($\gamma$ is the Sommerfeld coefficient.$\Theta_D$ and $\Theta_E$ are Debye and Einstein, temperatures respectively with $d$ the number of optical phonon modes. With n the number of atom per unit-cell = 9.18, and R being the gas constant, fit on experimental data yielded to $\Theta_D$ = 233 K,  $\Theta_E$ = 498 K and d = 0.13.

From the three fits, a large Sommerfeld coefficient $\gamma$ between 60.1 mJ mol$^{-1}$ K$^{-2}$ (cubic fit from eq. \ref{T3}), 65.7 mJ mol$^{-1}$ K$^{-2}$ (Debye-Einstein fit, see eq. \ref{debye}) and 88.5  mJ mol$^{-1}$ K$^{-2}$ from linear regression are extracted. Those large values tend to predict a heavy fermion behavior for this compound (see, discussion below). The heat capacity at room temperature is in the order of the estimation from Dulong-Petit law as observed in the inset of fig. \ref{fig:fig2}(d) with C$_p$ $\approx$ 228 J mol$^{-1}$ K$^{-1}$. For the rest of the discussion the low-temperature fit values will be considered as it proves to be more reliable.

\begin{figure*}
\centering
\includegraphics[width=0.5\textwidth]{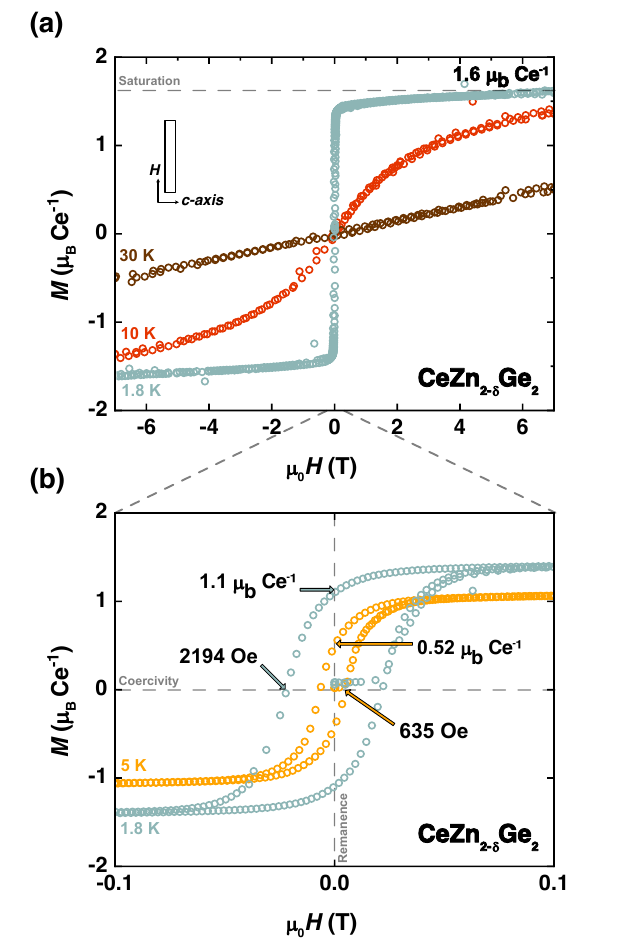}
\caption{(a) Field dependence of the in-\textit{ab}-plane magnetic moment per Ce for \ce{CeZn$_{2-\delta}$Ge2} measured between -7 T and 7 T at 1.8 K, 10 K and 30 K. An hysteresis loop is observed for temperature below $T_{\rm C}$ = 6.6 K. (b) Closer look at the hysteresis loops at $T$ = 1.8 and 5 K showing the coercivity and remanent field values.}
\label{fig:fig3}
\end{figure*}

Magnetization curves as a function of field at different temperatures are presented in fig. \ref{fig:fig3}. On \ref{fig:fig3}(b), at $T$ = 1.8 K and 5 K we observe a clear ferromagnetic behaviour displaying a hysteresis loop. A saturation moment of 1.6 $\rm \mu_B$ Ce$^{-1}$ is found above 4 T. The saturation is lower than the expected 2.14 $\rm \mu_B$ Ce$^{-1}$ for Ce$^{3+}$, but this value is in good agreement with the low saturation values reported for cerium 4f$^1$ ferromagnets, where typically even smaller values are observed. For example, values of 1.2 $\rm \mu_B$ Ce$^{-1}$ for \ce{CeRuPO}, 1 $\rm \mu_B$ Ce$^{-1}$ for \ce{CeAuGe}, and 0.4 $\rm \mu_B$ Ce$^{-1}$ for \ce{CeRh3B2} were observed. Again, this is in good agreement with the well-known quenching effect caused by the crystal field effect.\cite{wallace_rare_2012}

At 10 K, a residual ferromagnetic component is seen without a hysteresis while at 30 K, it has completely disappeared. There, the linear magnetization corresponds to the expected behavior of a paramagnetic state. In fig. \ref{fig:fig3} we show a zoom-in of the hysteresis loops at $T$= 1.8 K and 5 K, evidencing the ferromagnetic ordering. At 1.8 K and 5 K, coercivity field of 1270 Oe and 406 Oe and remanent moments of 0.4873 $\rm \mu_B$ Ce$^{-1}$ and 0.2003 $\rm \mu_B$ Ce$^{-1}$ are found, respectively. 

\subsection{Electronic transport properties}

\begin{figure*}
\centering
\includegraphics[width=0.5\textwidth]{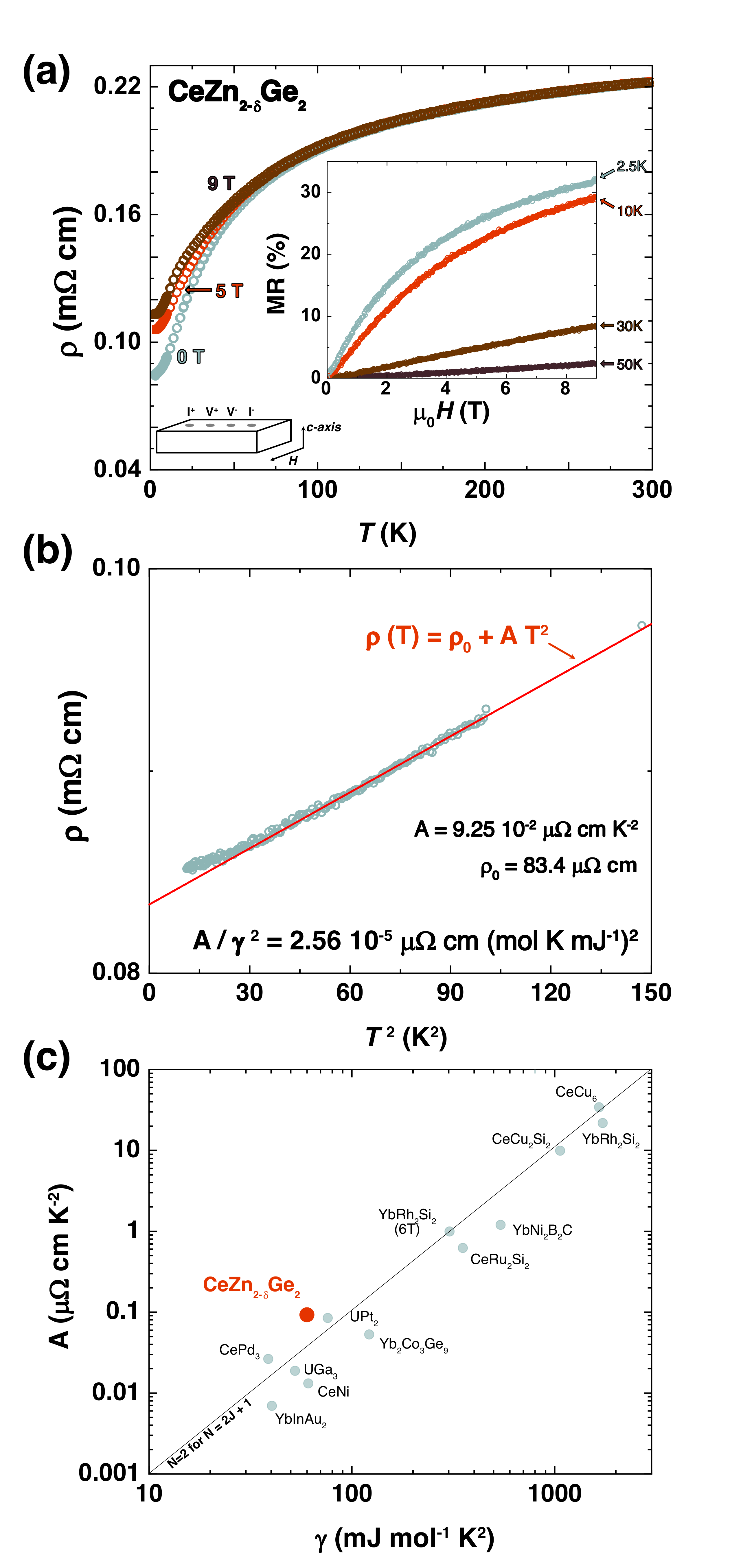}
\caption{(a) In-\textit{ab}-plane resistivity measured from 2K to 300 K under fields of 0 T, 5 T and 9 T. (Inset) Magnetoresistance of \ce{CeZn$_{2-\delta}$Ge2} measuring the dependence of resistivity \textit{vs.} field at different temperatures. (b)In-\textit{ab}-plane resistivity at low temperature, quadratic fit and the extracted Kadowaki-Woods ratio. (c) A coefficient vs Sommerfeld coefficient $\gamma$ for different heavy fermion metals. Our \ce{CeZn$_{2-\delta}$Ge2} is right next to the N = 2 line. Data extracted from reference \citenum{coleman_heavy_2007}.}
\label{fig:fig4}
\end{figure*}

In fig. \ref{fig:fig4} field- and temperature-dependent of in-\textit{ab}-plane resistivity measurements are presented in fields $\mu_0 H$ = 0 T, 5 T and 9 T in a temperature range between $T$ = 2 K and 300 K. \ce{CeZn$_{2-\delta}$Ge2} displays metallic behavior. Its resistivity in zero-field is in the order of the tenth of m$\Omega$ cm. It decreases from 0.22 m$\Omega$ cm at room temperature to 0.08 m$\Omega$ cm at $T$ = 2 K. The resistivity decreases and saturates below 20 K. A closer look at \ref{fig:fig4}b shows a deviation from the linearity of the above-$T_\textrm{C}$ resistivity below the transition temperature, which could link the electrical transport properties to the ferromagnetism below the transition. As atomic exchange disorder within the crystal structure was excluded from neutron diffraction while zinc deficiency was confirmed. The resistivity curve evolution then resembles the one from Kondo-lattice \ce{CeRuOP} \cite{krellner_cerupo_2007}, with a resistivity drop below 100 K, but the atomic disorder enhances scattering as well as the residual resistivity $\rho_0$. The temperature-dependent resistivity of \ce{CeZn$_{2-\delta}$Ge2} in zero field follows the classical behavior of heavy fermion compound: \textit{i.e.} at high temperatures we find a large saturated resistivity, while upon lowering the temperature only a moderate, very broad decrease in the resistivity is observed. This is caused by scattering induced by the coherency in the spin-flipping of conduction electrons off the local f-electrons (compare reference \citenum{coleman_heavy_2007}). While such a behavior is observed in Kondo-linked scattering, the resistivity minimum is not observed for this compound.

A positive magnetoresistance (MR) is observed below 100 K, with a stabilized maximum MR slightly above 30\% between 11 K and 2 K. In-\textit{ab}-plane MR curves at 2.5 K, 10 K, 30 K and 50 K are presented in the inset of fig. \ref{fig:fig4}(a). The MR is determined according to

\begin{equation}
\mathrm{MR = [\rho(\mu_0 \textit{H}) - \rho(0T)] / \rho(0T) \times 100}
\end{equation}

A positive magnetoresistance begins to be displayed from 50 K down to 1.8 K. A peak of 32 \% is reached at 2.5 K for 9 T, while at 10 K for the same field, the magnetoresistance is approximately 28\%. Few materials with large magnetoresistance are found in different compounds of the \ce{ThCr2Si2} and \ce{CaBe2Ge2} families including the antiferromagnet \ce{PrPd2Si2} with a positive MR of 20\%,\cite{anand_magnetic_2007} \ce{SmMn2Ge2} with -16\% at 125 K\cite{brabers_giant_1993} or \ce{LaMn2Ge2} which exhibits a positive MR of about 100\% at low temperatures.\cite{williams_synthesis_2003,mallik_large_1997} 

What is surprising is indeed a positive MR for a ferromagnetic ordering with a low amount of reported compounds.  Ferromagnetic \ce{SmPd2Ga2} is for example found with a positive 100\% MR. Further studies will be needed to elucidate on the interplay between the ferromagnetic state and the positive magnetoresistance. Though it was previously reported that heavy-electron/fermion systems can show large positive magnetoresistance at low enough temperatures and with our large calculated Sommerfeld-coefficient, this could be an explanation.\cite{remenyi_magnetoresistivity_1983,rauchschwalbe_magnetic_1985,sumiyama_low_1985}. The fact that the magnetoresistance at 2.5 K and 10 K are of similar scale, and that the MR starts at temperature higher than the ferromagnetic transition reinforce the idea that the MR is not linked to the ferromagnetic state. The MR could have multiple origins including the zinc deficiency, Kondo effect, and heavy-fermionic properties. Additional measurements are required to comprehensively understand the magnetic structure and its link to the electronic properties.

\subsection{Heavy fermion compound}

The observation of indications for Kondo scattering in the temperature-dependent resistivity and the large Sommerfeld coefficient $\gamma$ in the specific heat (discussed above) hint towards \ce{CeZn$_{2-\delta}$Ge2} being a heavy fermion compound. At low temperatures and above the ferromagnetic transition, we obtain for the resistivity $\rho$(T) a quadratic temperature dependence. In the inset of fig. \ref{fig:fig4}(b), the resistivity $\rho$(T) is plotted versus the temperature squared T$^2$, manifesting the quadratic temperature dependence. In the low-temperature region between 6 K and 10 K, the resistivity $\rho$(T) can be well-fitted with

\begin{equation}
\mathrm{\rho(\textit{T}) = \rho_0 + A \times \textit{T}^2}   
\end{equation}

Here, A is equal to 9.25 $\times$ 10$^{-2}$ $\rm \micro \Omega$ cm K$^{-2}$. This quadratic temperature dependence of the resistivity indicates the formation of the Fermi-liquid state. The Kadowaki-Woods ratio $\rm {A}/\gamma^2$ is a measure of the magnitude of electron-electron correlation \cite{rice_electron-electron_1968,kadowaki_universal_1986} and is determined using $\gamma$ from the specific heat C$_P$ measurements. Here, we obtain a ratio of $\rm {A}/\gamma^2$ = 2.56 $\times$ 10$^{-5}$ $\rm \micro \Omega$ cm (mol K mJ$^{-1}$)$^2$. This value is about 30 times larger than for a normal metal, which is an indication for strong electron-electron correlation in this material (compare, \textit{e.g.} references \citenum{von_rohr_superconductivity_2014} and \citenum{jacko_unified_2009}). In fig. \ref{fig:fig4}(c) we show the A coefficient vs $\gamma$ Sommerfeld coefficient plot for spin degeneracy N = 2J + 1 of the magnetic ions with J = 1/2, \textit{e.g.} Ce$^{3+}$ 4f$^1$ (see reference \citenum{coleman_heavy_2007}). The material at hand, \ce{CeZn$_{2-\delta}$Ge2}, lies right on this N = 2 curve, next to \ce{UPt2}. This observation is strong evidence for \ce{CeZn$_{2-\delta}$Ge2} to be a heavy fermion compound.

We, furthermore, estimate the Wilson ratio, which is another measure for electron-electron correlation.\cite{wilson_renormalization_1975} From the temperature independent contribution to the susceptibility, we estimate the Pauli paramagnetic contribution of $\chi_0 = 1.01 \times 10^{-3}$. In heavy-electron systems, the Wilson ratio W = ($\pi^2 {k_B}^2 / {\mu_B}^2) \chi_0/\gamma$ (with $k_B$ being the Boltzmann constant and $\gamma$ the Sommerfeld coefficient) is close to one.\cite{nozieres_fermi-liquid_1974,nozieres_kondo_1980} Our calculation estimates the Wilson ratio to be close to 1.31, which is a further strong indication for the incompressible and local character of heavy fermions.

\section{Conclusion}
The new ternary compound \ce{CeZn$_{2-\delta}$Ge2} in the cerium-germanium-zinc ternary system has been reported. Large single crystals of it have been synthesized by using Zn/In flux. The crystal structure of this compound has been determined by means of SXRD. It crystallizes in the \textit{P}4/\textit{nmm} space-group with cell parameters $a$ = 4.20635(16) $\si{\angstrom}$ and $c$ = 10.3338(7) $\si{\angstrom}$, with a structure that is isostructural to the \ce{CaBe2Ge2}-type structures. The refined composition for the collected crystal is \ce{CeZn$_{1.591(11)}$Ge2}, the structure is ordered and zinc deficient as confirmed by means of preliminary powder neutron diffraction. \ce{CeZn$_{2-\delta}$Ge2} shows metallic behavior on the whole measured temperature range between $T$ = 2 K and 300 K. For large fields, below 10 K a large positive magnetoresistance of MR slightly above 30 \% is observed. The compound is one of the rare examples of a Ce-based ferromagnet. We have determined its transition temperature to be $T_{\textrm{C}}$ = 6.6 K using an Arrott plot and specific heat measurement. We find \ce{CeZn$_{2-\delta}$Ge2} to be a heavy fermion compound as evidenced by the analysis of the Kadowaki-Wood relations, the Wilson ratio, and the indications for Kondo scattering in the resistivity.

We expect the results at hand to generally motivate significant additional studies into this material, as \ce{CeZn$_{2-\delta}$Ge2} is a versatile and promising model platform for the investigation of Ce-ferromagnetism and strong electron-electron correlations coupled with complex crystal chemistry.

\section{Experimental}
All manipulations of the starting materials were performed under inert conditions. The elements, Ce (powder $\approx$ 40 mesh, 99.9\%), Ge (powder $\approx$ 100 mesh, 99.999 \%), In (beads, 2-5 mm, 99.999\%), and Zn (powder $\le$ 150 $\rm \mu$m, 99.995 \%), all from Aldrich, were used as received. \ce{CeZn$_{2-\delta}$Ge2} crystals were synthesized using a starting ratio of Ce:1;Ge:2;Zn:7.2;In:4.8 for a total mass of 8 g. All the elements were placed in a crucible with quartz wool on top in a silica tube. The tube was sealed under $\approx$ 300 mbar of argon. The tube was placed in a muffle furnace and the temperature was raised to 1373 K in 24 hours and kept at that temperature for 24 hours. The crystals were grown by lowering the temperature slowly to 773 K over a duration of 19 days. The flux was then centrifuged away. 

SXRD analyses were collected at 160(1) K using Cu K$\alpha$ radiation ($\lambda$ = 1.54184 $\si{\angstrom}$) on a XtaLAB Synergy, Dualflex, Pilatus 200K. Pre-experiment, data collection, manual data reduction, and sphere absorption correction were carried out with the program suite CrysAlisPro.\cite{noauthor_crysalispro_2014} Using Olex2 crystallography software,\cite{dolomanov_olex2_2009} the structure was solved with the SHELXT Intrinsic Phasing solution program \cite{sheldrick_shelxt_2015} and refined with the SHELXL 2018/3 program \cite{sheldrick_shelxl_2015} by full-matrix least-squares minimization on F${^2}$.  Energy dispersive X-ray spectroscopy analyse was also performed to check the composition using a scanning electron microscope (SEM; ZEISS Gemini 450, 15 kV). The samples purity was investigated by means of powder X-ray diffraction (PXRD) measurements on a STOE STADIP diffractometer with Cu K${\alpha_1}$ radiation ($\lambda$  = 1.540600 Å). Powder neutron diffraction was realized on the High-Resolution Powder Diffractometer for Thermal Neutrons (HRPT) beamline at PSI, Villigen, Switzerland using a 1.886 $\si{\angstrom}$ wavelength at temperature of $T$ = 175 K.

The in-\textit{ab}-plane magnetic susceptibility was measured using a SQUID MPMS3 (Quantum Design) equipped with the vibrating sample magnetometer (VSM) option. The measurements were performed on a 2.9 mg single crystal in a temperature range between $T$ = 1.8 to 300K in sweep mode at a 1 K min$^{-1}$ rate below 10 K and 5 K min$^{-1}$ rate above. Using a field of $\rm \mu_0 H$ = 10 mT, the susceptibility was measured with a field-cooled procedure. The in-\textit{ab}-plane magnetization was measured from -7 T to 7 T at 1.8 K, 5 K, 10 K and 30 K. Fields below $\pm$4000 Oe were incremented at a 5 Oe sec$^{-1}$ rate while absolute values above were incremented at a 200 Oe sec$^{-1}$, and points were measured in continuous mode. Arrott-plot using in-\textit{ab}-plane magnetization measurements was created between 6.4 K and 7 K at 0.1 K intervals with fields up to 0.1 T with an increase rate of 200 Oe sec$^{-1}$.

Low-temperature in-\textit{ab}-plane resistivity and magnetoresistance were measured using a Quantum Design Physical Properties Measurement System (PPMS). The resistivity was measured from 2 K to 300 K in sweep mode at 0.1 K min$^{-1}$ below 10K and at 2 K min$^{-1}$ above, at 0 T, 5 T and 9 T. Magnetoresistance was measured at 2.5 K, 10 K, 30 K and 50 K with fields from 0 T to 9 T. For the measurements, 4 probe electrical contacts were made using conductive mixed silver paste and connected to the puck using gold wires. The same device was used to measure specific heat capacity from 2 K to 300 K with logarithmic measurement scale, the commercial Quantum Design specific heat option with a relaxation technique was used.

\begin{acknowledgement}

This  work  was  supported  by the Forschungskredit No. FK-20-117, the Ernst-G\"ohner Foundation, and the Swiss National Science Foundation under Grant No. PCEFP2\_194183. We thank Dorota Walicka, Dr. Nami Matsubara, Dr. Alejandro Jimenez Muñoz, Miguel Ángel Jiménez Herrera, Dr. Vladimir Pomjakushin and Prof. Dr. Marc Janoschek for helpful discussions.\\

\end{acknowledgement}

\section{Competing interests}

The authors have no competing interests to declare.

\begin{suppinfo}
Additional data can be found in supplementary information including powder X-ray and neutron diffraction refinements, EDS measurement, 1 T magnetic susceptibility, 10 mT derivative susceptibility and the heat capacity over $T$ \textit{vs} $T^2$ linear fit.
\end{suppinfo}

\bibliography{ref}

\end{document}


\begin{abstract}
\end{abstract}


\begin{figure*}
\centering
\includegraphics[width=\textwidth]{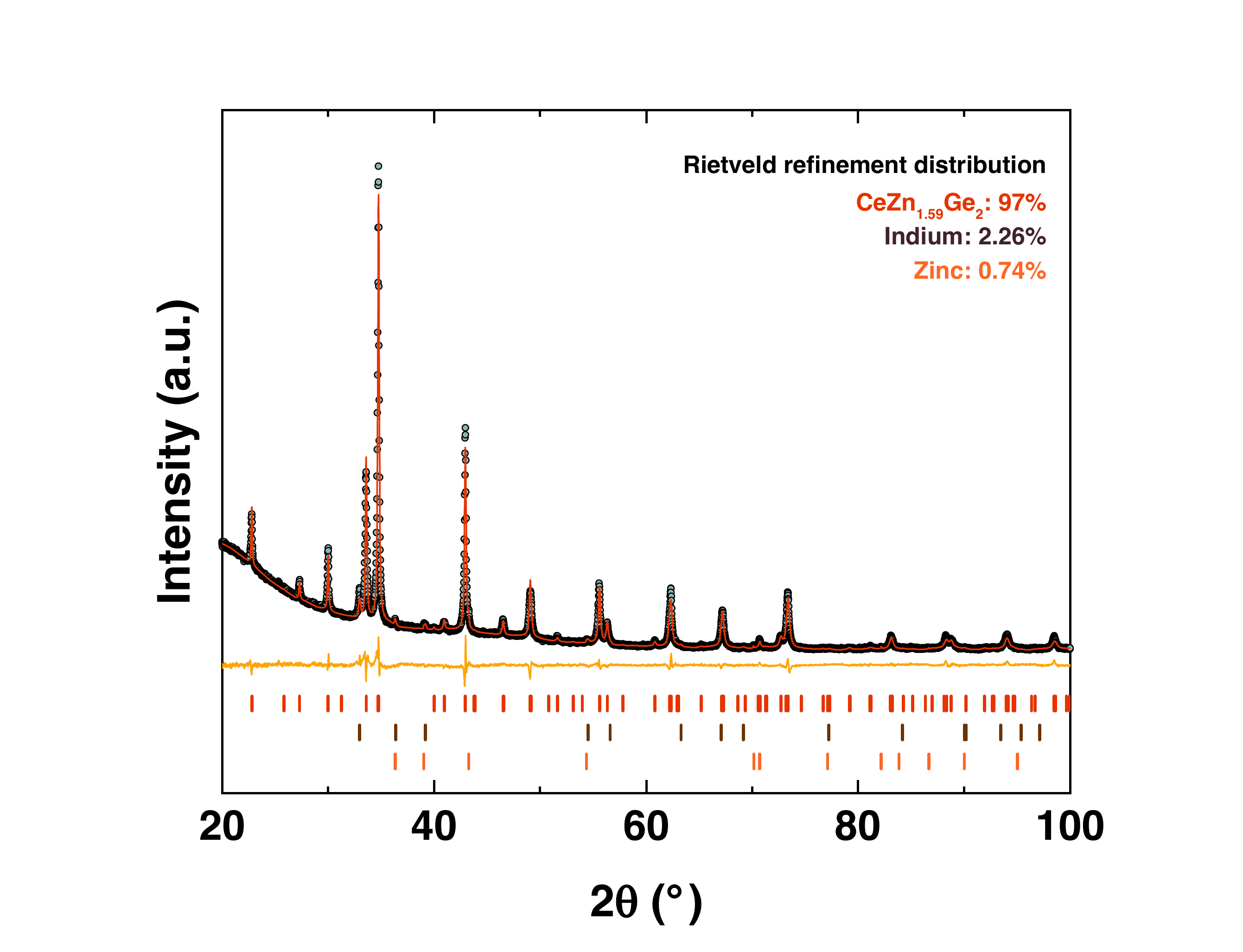}
\caption{PXRD  pattern  and  Rietveld  refinement  of crushed crystals of \ce{CeZn_{2-\delta}Ge2}. The  plot  is  represented  as  follows:   observed(colored dots),  calculated (red line) and difference (colored line) intensities.  The Bragg positions of the phases are indicated with colored vertical bars.}
\label{Sfig1}
\end{figure*}

\begin{figure*}
\centering
\includegraphics[width=\textwidth]{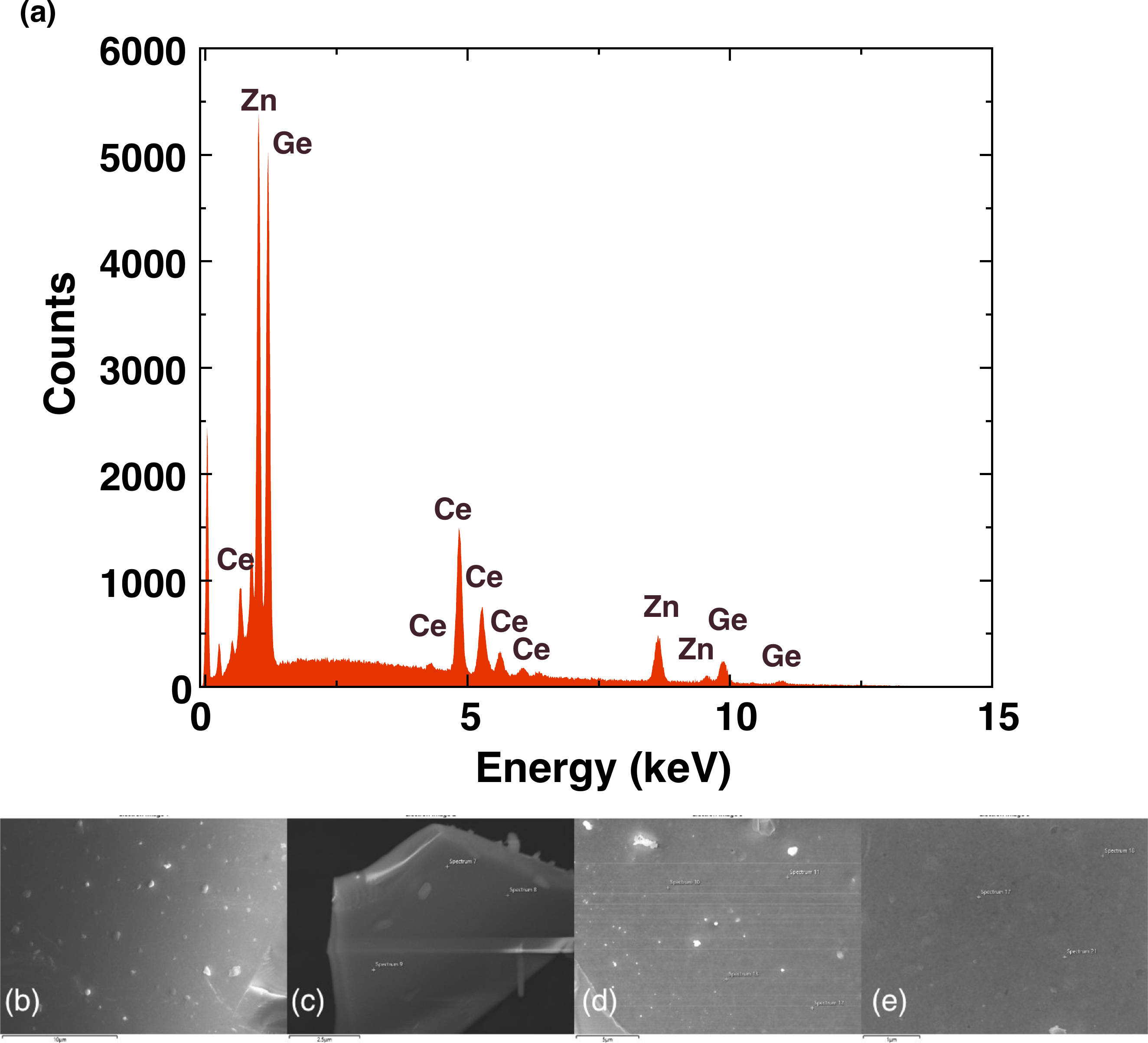}
\caption{EDS measurement on a crystal of \ce{CeZn_{2-\delta}Ge2} (a) EDS spectrum showing Ce, Ge, and Zn atoms energy peaks, (b) (c) (d) and (e) are SEM pictures made on four different parts used for averaging the atomic compositions.}
\label{Sfig2}
\end{figure*}

\begin{table*}
\centering
\caption{EDS atomic ratio measured on different spectra and averaging values on a \ce{CeZn_{2-\delta}Ge2}}
\label{STable1}
\resizebox{0.5\textwidth}{!}{%
\begin{tabular}{lllll}
\hline
\textbf{Spectrum   Label} & \textbf{Zn} & \textbf{Ge} & \textbf{Ce} & \textbf{Total} \\
Spectrum   1 & 36.37 & 40.58 & 23.05 & 100.00 \\
Spectrum   2 & 35.53 & 41.24 & 23.23 & 100.00 \\
Spectrum   3 & 34.44 & 42.56 & 23.00 & 100.00 \\
Spectrum   4 & 35.59 & 41.92 & 22.49 & 100.00 \\
Spectrum   5 & 35.43 & 40.89 & 23.68 & 100.00 \\
Spectrum   6 & 37.75 & 40.53 & 21.71 & 100.00 \\
Spectrum   7 & 33.18 & 44.11 & 22.71 & 100.00 \\
Spectrum   8 & 36.07 & 40.72 & 23.21 & 100.00 \\
Spectrum   9 & 34.85 & 42.50 & 22.66 & 100.00 \\
Spectrum   10 & 34.93 & 41.50 & 23.57 & 100.00 \\
Spectrum   11 & 34.03 & 42.88 & 23.09 & 100.00 \\
Spectrum   12 & 34.61 & 42.70 & 22.68 & 100.00 \\
Spectrum   13 & 35.18 & 41.51 & 23.31 & 100.00 \\
Spectrum   17 & 36.03 & 42.00 & 21.98 & 100.00 \\
Spectrum   18 & 36.42 & 41.78 & 21.81 & 100.00 \\
Spectrum   21 & 36.15 & 41.89 & 21.96 & 100.00 \\
 &  &  &  &  \\
 \hline
\textbf{Statistic} & \textbf{Zn} & \textbf{Ge} & \textbf{Ce} & \textbf{} \\
Max & 37.75 & 44.11 & 23.68 &  \\
Min & 33.18 & 40.53 & 21.71 &  \\
Average & 35.41 & 41.83 & 22.76 &  \\
Standard   Deviation & 1.09 & 0.97 & 0.62 & 
\end{tabular}%
}
\end{table*}

\begin{figure*}
\centering
\includegraphics[width=\textwidth]{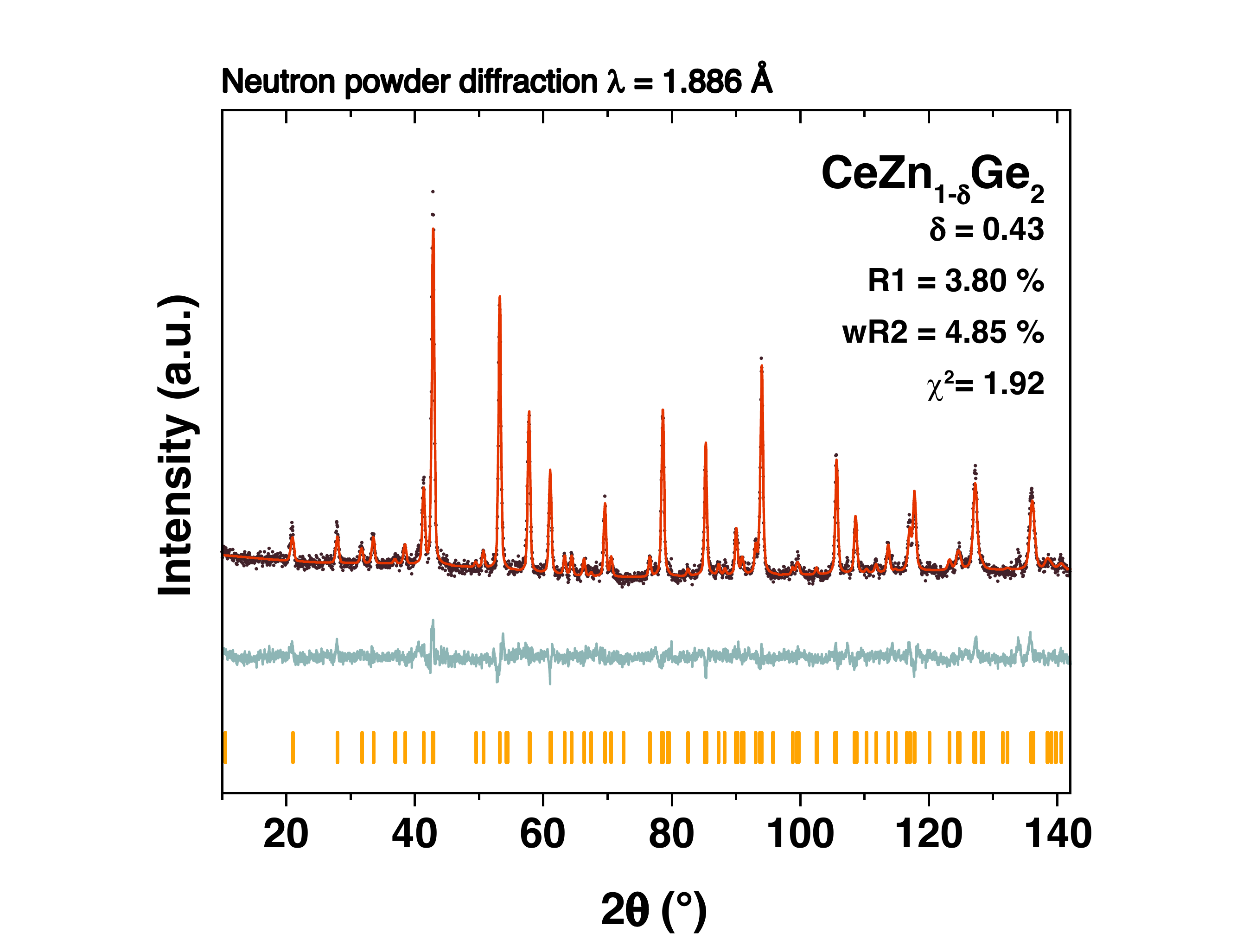}
\caption{Rietveld refinement plot from powder neutron diffraction using the same model as obtained from single X-ray diffraction. Observed, calculated and difference intensities are represented by brown dots, red line and blue line. Bragg positions are represented by yellow bars.}
\label{Sfig3}
\end{figure*}

\begin{figure*}
\centering
\includegraphics[width=\textwidth]{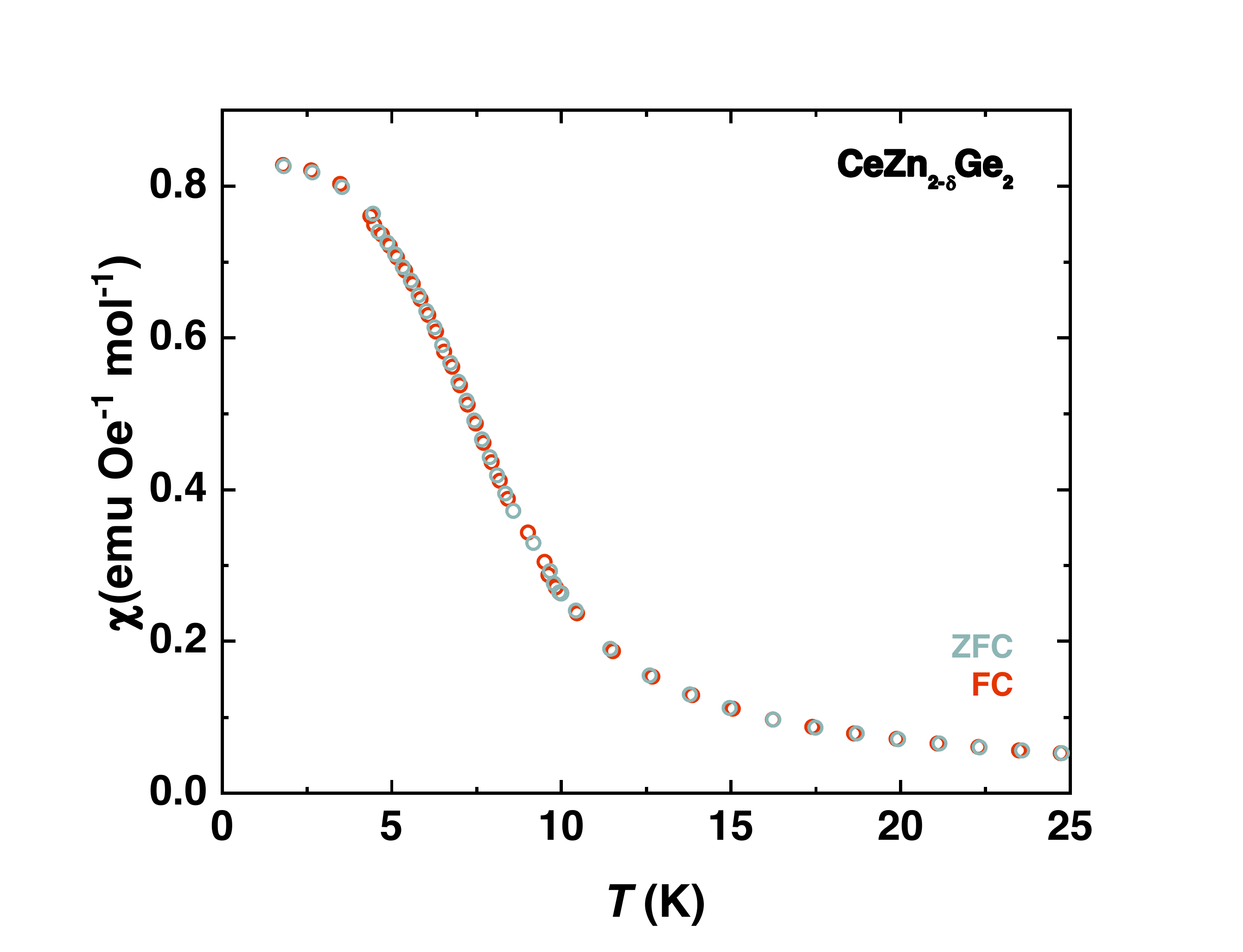}
\caption{Zero-field-cooled and field-cooled of the in-plane susceptibility of \ce{CeZn_{2-\delta}Ge2} under $\mu_0 H$ = 1T.}
\label{Sfig4}
\end{figure*}

\begin{figure*}
\centering
\includegraphics[width=\textwidth]{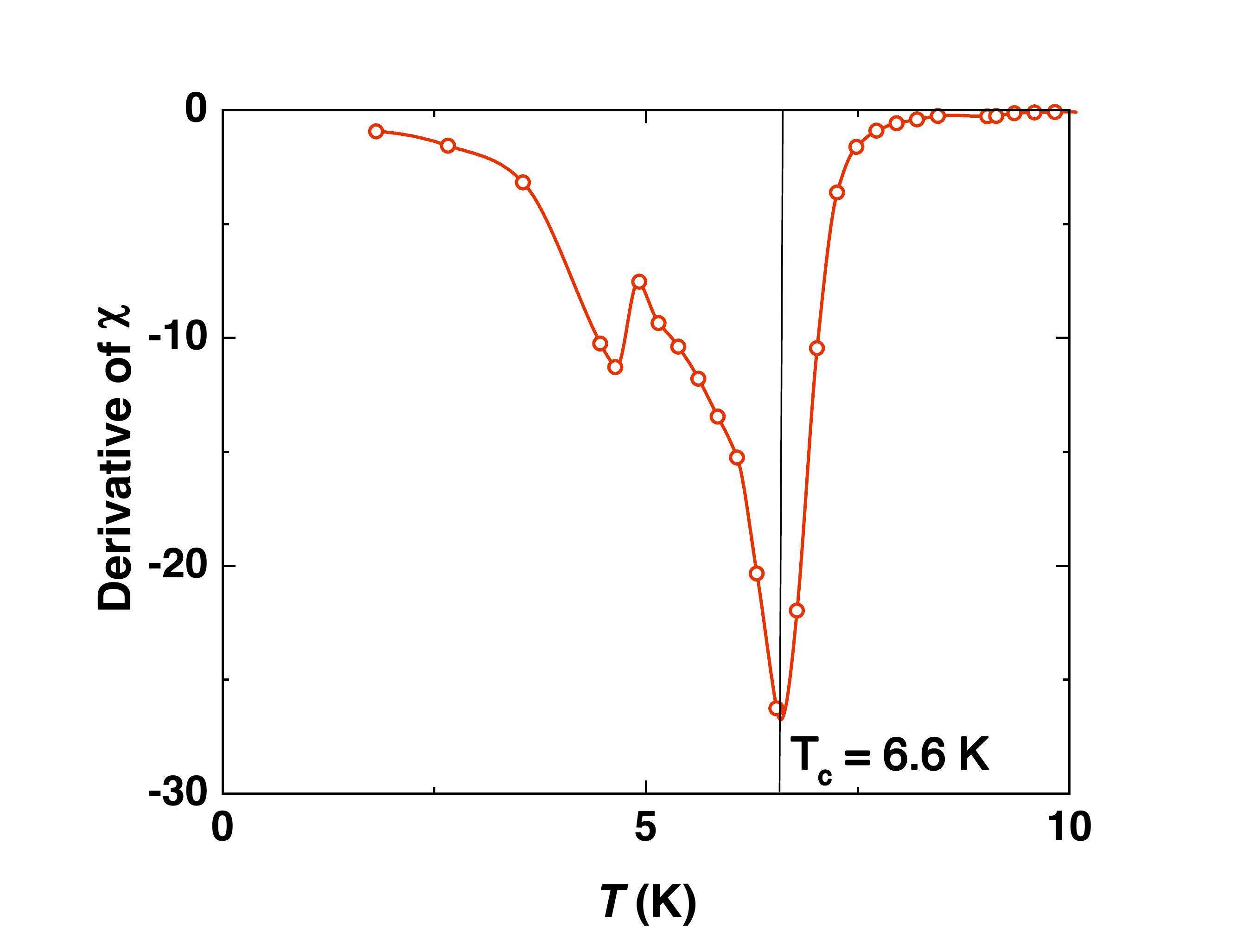}
\caption{Derivative curve of the in-plane magnetic susceptibility under $\mu_0 H$ = 10 mT of \ce{CeZn_{2-\delta}Ge2} showing the maximum of the inflation at 6.6 K.}
\label{Sfig5}
\end{figure*}

\begin{figure*}
\centering
\includegraphics[width=\textwidth]{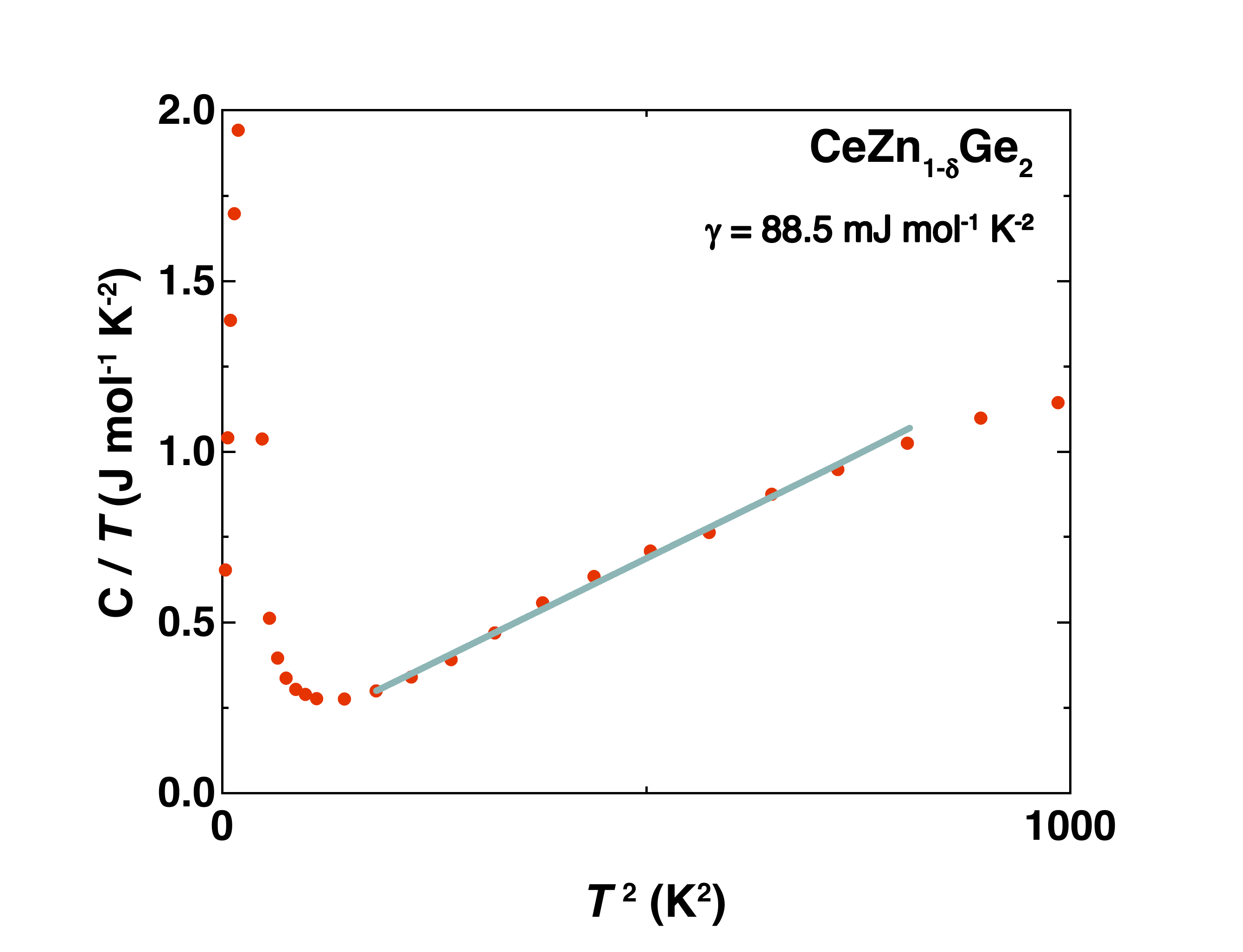}
\caption{C/\textit{T} \textit{vs.} T$^2$ between 35 K and 2 K and its linear regression above the transition.}
\label{Sfig6}
\end{figure*}